\documentclass[12pt,preprint]{aastex}

\shorttitle{X-wind Transport}
\shortauthors{Renyu Hu}

\begin{document}

\title{Transport of First Rocks of The Solar System by X-winds}

\author{Renyu Hu}
\affil{Department of Earth, Atmospheric and Planetary Sciences, Massachusetts Institute of Technology, Cambridge, MA 02139}
\email{hury@mit.edu}

\begin{abstract}
It has been suggested that chondrules and calcium-aluminum-rich inclusions (CAIs) were formed at the inner edge of the protoplanetary disk and then entrained in magnetocentrifugal X-winds. We study trajectories of such solid bodies with the consideration of the central star gravity, the protoplanetary disk gravity, and the gas drag of the wind. The efficiency of the gas drag depends on a parameter $\eta$, which is the product of the solid body size and density. We find that the gravity of the protoplanetary disk has a non-negligible effect on the trajectories. If a solid body re-enters the flared disk, the re-entering radius depends on the stellar magnetic dipole moment, the disk's gravity, the parameter $\eta$, and the initial launching angle. The disk's gravity can make the re-entering radius lower by up to 30\%. We find a threshold $\eta$, denoted as $\eta_t$, for any particular configuration of the X-wind, below which the solid bodies will be expelled from the planetary system. $\eta_t$ sensitively depends on the initial launching angle, and also depends on the mass of the disk. Only the solid bodies with a $\eta$ larger than but very close to $\eta_t$ can be launched to a re-entering radius larger than 1 AU. This size-sorting effect may explain why chondrules come with a narrow range of sizes within each chondritic class. In general, the size distributions of CAIs and chondrules in chondrites can be determined from the initial size distribution as well as the distribution over the initial launching angle.
\end{abstract}


\section{Introduction}

X-winds are high-speed bipolar collimated jets around young stellar objects (YSOs) powered by enhanced magnetic activities and disk-magnetosphere interactions (Shu et al. 1996, 1997). An X-wind can be driven magnetocentrifugally from the inner edge of the disk where accreting gas is either diverted onto stellar field lines to flow onto the star or to be flung outwards with the wind (Shu et al. 1994a,b).

X-winds are also suggested to be the underlying mechanism for the production of calcium-aluminum-rich inclusions (CAIs) and chondrules for our Solar system (Shu et al. 2001). The subsolar reconnection ring (X-region) at the inner edge of the protoplanetary disk provides high-temperature environment to form CAIs and chondrules, and the X-winds can lift the processed rocks back to colder regions of the disk. At least 50\% by mass of all the magnesium-iron silicates that finally constructed planetesimals and planets may have been processed as chondrule or CAIs (Tayler 1992; Shu et al. 1996).

Isotopic measurements also provide evidence of the X-wind origin of planetary materials. CAIs have $^{18}$O/$^{16}$O and $^{17}$O/$^{16}$O ratios that are several percent lower than any other meteoritic and planetary materials in the Solar system (Clayton 1993). This isotopic feature can be explained by photochemical self-shielding of carbon monoxide in the solar nebula, which requires that planetary materials passed through the inner part of the accreting protoplanetary disk (Clayton 2002). Again the processed materials need to be returned back to colder regions by X-winds.

However, there are also other formation mechanisms that have been proposed to explain the chemical, mineral and isotopical properties of chondrules, such as exothermic chemical reactions, nebular lightning, magnetic reconnection flares, gas dynamic shock waves in the protoplanetary nebula, radiative heating, etc. (see Rubin 2000 for an extensive review). Desch et al. (2010) critically examines the X-wind model for the formation of chondrules and CAIs against isotopical and mineralogical evidences, but also advocates detail modeling on the fate of the particles launched by the outflow to see if the particles can be retained in proper annuli of the disk.

The parent bodies of chondritic meteorites originate mostly from the asteroid belt at about 2.5 astronomical units (AU) from the Sun. Recent observations find evidence of CAI-type and chondrulelike particles present in a short-term comet 81P/Wild 2 (Swindle \& Campins 2004; Brownlee et al. 2006; Flynn et al. 2006; Nakamura et al. 2008). Wooden (2008) concluded from mineralogy investigations that these refractory grains should be pre-accretionary with respect to the formation of asteroids and efficient outward radial transport mechanism is required. This discovery requires large-scale radial transport mechanisms, such as X-winds and the turbulent diffusion of particles (e.g. Ciesla 2009). However, it has not been studied whether X-winds are able to drive solid bodies to the radius of the Kuiper belt, or $>$30 AU.

It has not yet been clearly explained under what conditions X-wind can deliver chondrules and CAIs to the asteroid belt or the Kuiper belt. Shu et al. (1996) and Shang (1998) provides several trajectories of solid bodies in X-winds in the dimensionless form without the consideration of protoplanetary disk gravity nor the vertical geometry of the disk. Liffman (2005) investigated the size-sorting effect of chondrules from magnetic pressure driven jet flows, but only for distance $\le$3AU from the Sun. In this paper we study the transport of solid bodies (e.g. CAIs and chondrules) by the X-winds, with the protoplanetary disk's gravity and vertical geometry taken into account. We also investigate whether and where the launched solid bodies re-enter the disk and how the re-entering radius depends on the stellar magnetic dipole moment and the size and density of the solid bodies.

\section{Model}

\subsection{X-wind Configurations}

In this section we provide an introduction on the X-wind configurations based on Shu et al. (1994a, 1996), and identify several stellar and disk parameters that govern the X-wind.

An accreting protoplanetary disk that interacts with the rotating magnetosphere of a protostar with a pure dipole is truncated at a radius of
\begin{equation}
R_{\rm X}=C_1\bigg(\frac{\mu_*^4}{GM_*\dot{M}_{\rm D}^2}\bigg)^{1/7}\ ,\label{RX}
\end{equation}
where $\mu_*$ is the stellar dipole moment, $M_*$ is the stellar mass, $\dot{M}_{\rm D}$ is the inflow rate of the disk, $G$ is the gravitational constant, and $C_1$ is a dimensionless coefficient that we assume to be unity in the following (e.g. Shu et al. 1996).
In steady state, the inner disk edge corotates with the star at the Keplerian angular velocity of
\begin{equation}
\Omega_*=\Omega_{\rm X}=\bigg(\frac{GM_*}{R_{\rm X}^3}\bigg)^{1/2}\ .\label{Omega}
\end{equation}

The wind has a terminal velocity $\bar{v}_w$ when leaving the X region. The terminal velocity can be written as
\begin{equation}
\bar{v}_w=(2J_g-3)^{1/3}R_{\rm X}\Omega_{\rm X}\ ,\label{vw}
\end{equation}
where $J_g$ is the dimensionless specific angular momentum of the gas (Shu et al. 1994).

The formation of CAIs requires the thermal treatment at 1800$\sim$2200 K. CAIs and chondrules entrained in X-wind experience the peak temperature when they are lifted into the direct sunlight. The peak temperature is given by Shu et al. (1996) as
\begin{equation}
T_{\rm peak}=\bigg\{\frac{L_*}{16\pi\sigma R_{\rm X}^2}+\frac{L_*}{8\pi^2\sigma R_*^2}\bigg[\arcsin\bigg(\frac{R_*}{R_{\rm X}}\bigg)-\bigg(\frac{R_*}{R_{\rm X}}\bigg)\bigg(1-\frac{R_*^2}{R_{\rm X}^2}\bigg)^{1/2}\bigg]\bigg\}^{1/4}\ ,\label{Tpeak}
\end{equation}
where $L_*$ is the stellar luminosity, $\sigma$ is the Stefan-Boltzmann constant, and $R_*$ is the radius of the star.

From above we see that the physical properties of X-winds are determined by three parameters of a planetary system: $M_*$, $\mu_*$ and $\dot{M}_{\rm D}$. In theory, the X-wind can exist in either the `embedded' stage, when the protostar is still embedded in its natal envelope of gas and dust; or the `revealed' stage, when the outflowing wind has reversed the infall of the envelope and revealed the central star and the accretion disk (Shu et al. 1996). At different stages, the star has different masses, the stellar magnetic dipole moment varies in different ranges, and the accretion rate has different values. As long as the lifetime of these two stages are fairly well constrained by the observations of young stellar objects, we use typical values of $M_*$ and $\dot{M}_{\rm D}$ for each stages, and let $\mu_*$ vary in different ranges, respectively. Following Shu et al. (1997), we define a high state, an average state and a low state according the value of $\mu_*$ for each stage, as shown in Table \ref{Tab1}. Key X-wind properties $R_{\rm X}$, $\Omega_{\rm X}$, $\bar{v}_w$ and $T_{\rm peak}$ of each state are also given in Table \ref{Tab1}. As evident from Table \ref{Tab1}, a protosun with stronger magnetic field truncates the protoplanetary disk at a larger radius, and therefore has lower $T_{\rm peak}$ and $\bar{v}_w$. Notably, CAIs can only form during the EA and EL cases.

For each state, the gas density and flow velocity of the X-wind can be determined numerically, from a set of equations involving the conservation of mass, specific momentum and specific energy (see Shang 1998 for the detail formulation). In the following, we formulate the equation of motion in the inertial frame, with the cylindrical coordinates $(l, \phi, z)$ centered at the star and with $z=0$ defining the midplane of the disk. The asymptotic behavior of the X-wind dynamics sufficiently far from the X-region (i.e. $l>50R_{\rm X}$) is: (1) the poloidal velocity reaches the terminal velocity; (2) the toroidal velocity decreases as $l^{-1}$; (3) the density of the wind falls off as $l^{-2}$. Assuming the streamlines to be straight lines originating from the X-region and the open angle of the bipolar wind to be $\theta_w=45^o$, we follow the procedure of Shang (1998) to construct the X-wind as the background for the motion of the solid body.

\subsection{Extended Model}

In this section we set up the framework to calculate the trajectories of solid bodies entrained in the X-wind. A solid body with size of $R_s$ and density of $\rho_s$ entrained in a X-wind is subject to three forces: the gravity of the star, the gravity of the protoplanetary disk, and the aerodynamic drag of the X-wind. The total force exerted on the solid body is
\begin{equation}
{\bf F}=-\frac{GM_*m_s}{r^3}{\bf r}-A(l,z)m_s{\bf e}_z-B(l,z)m_s{\bf e}_l+{\bf F}_{\rm drag}\ ,\label{EM1}
\end{equation}
where $m_s$ is the mass of the solid body, and $r$ is the distance from the solid body to the center of star. The second and the third terms in the right-hand side of equation (\ref{EM1}) represent the protoplanetary disk gravity, which has not been included in previous trajectory calculations. $A(l,z)$ and $B(l,z)$ can be calculated by the superposition of gravitational contributions from surface elements throughout the disk, viz.
\begin{eqnarray}
&&A(l,z)=-\frac{\partial}{\partial z}\int_{R_{\rm in}}^{R_{\rm out}}\int_0^{\pi} \frac{2G\Sigma(l')l'd\phi' dl'}
{\sqrt{l'^2+l^2+z^2-2l'l\cos\phi'}} \ ,\\
&&B(l,z)=-\frac{\partial}{\partial l}\int_{R_{\rm in}}^{R_{\rm out}}\int_0^{\pi} \frac{2G\Sigma(l')l'd\phi' dl'}
{\sqrt{l'^2+l^2+z^2-2l'l\cos\phi'}} \ ,
\end{eqnarray}
where $l'$ and $\phi'$ are dummy variables of integration spanning the whole disk, $R_{\rm in}$ and $R_{\rm out}$ are the inner and outer edge of the protoplanetary disk, and $\Sigma(l)$ is the surface density of the disk. In the calculation of disk's gravity, we assume the geometry of the disk to be a thin plane at $z=0$, and use $R_{\rm in}=R_{\rm X}$ and $R_{\rm out}=40$ AU. The integrals can be evaluate numerically once we know the surface density profile of the protoplanetary disk.

We adopt that the disk has the mass distribution as the standard minimum mass solar nebula (MMSN) given by Hayashi (1981). However, the density of the protoplanetary disk might be much higher. For example, Desch (2007) estimated a denser MMSN based on the NICE model. The disk's surface density in our model is parameterized as
\begin{equation}
\Sigma(l)=f\times1700 \bigg(\frac{l}{1\ {\rm AU}}\bigg)^{-3/2}\ {\rm g}\ {\rm cm^{-2}}\ ,\label{MMSN}
\end{equation}
where $f$ is a multiplying factor specifying the mass of the disk and the rest is the MMSN profile given by Hayashi (1981). Note that Desch (2007) suggested the surface density exponent to be $-2.2$ instead of $-1.5$, but in this work we adopt the profile by Hayashi (1981). The acceleration due to the disk gravity $A(l,z)$ and $B(l,z)$ can then be computed numerically.

The protoplanetary disk has the vertical extension that may intercept the solid body trajectory and influence the radius where the solid body re-enters the disk. Assuming the disk is supported by the thermal pressure, the scale height of the disk is
\begin{equation}
H(l)=\sqrt{\frac{2k_{\rm B}T(l)l^3}{GM_{*}\mu M_{\rm H}}}\ ,\label{DVert}
\end{equation}
where $k_{\rm B}$ is the Bolzmann constant, $\mu$ is the mean molecular mass in the atomic mass unit (for the solar abundance $\mu=2.3$), $M_{\rm H}$ is the mass of hydrogen atom, and $T(l)$ is assumed to be the temperature of the MMSN disk as $T(l)=280\ l^{-1/2}$ K (see Hartmann 2009). In the following we regard the scale height in equation (\ref{DVert}) as the `boundary' of the disk. Once the vertical position of the solid body becomes smaller than the scale height of the disk, $z<H$, the solid body re-enters the disk, and then the trajectory calculation is terminated. We do not intend to follow the trajectories of the solid bodies inside the disk. Once the horizontal position of the solid body becomes outside the disk, or $l>40$ AU, the solid body is ejected from the planetary system. In this work, we assume a thin disk when calculating the gravitational force from the disk. Still, we do consider the vertical geometry of the disk using a realistic "flared disk" model when determining the radius where the solids re-enter the disk. We will demonstrate that the vertical structure of the disk has a significant effect on the re-entering radius.

The Epstein gas drag applies when the relative velocity of the solid body and the wind is less than the sound speed of the wind, and the radius of the solid body is smaller than the mean free path of the gas. For example, the mean free path of the wind is $10\sim10^2$ cm during the embedded stage, and $10^2\sim10^3$ cm during the revealed stage. Therefore when calculating the trajectories of chondrule-or-CAI-size solid bodies entrained in X-winds, the Epstein gas drag law should be used, viz.
\begin{equation}
{\bf F}_{\rm drag}=\frac{4}{3} \pi R_s^2 \rho_w [({\bf v}_w-{\bf v}_s)^2+c_w^2]^{1/2} ({\bf v}_w-{\bf v}_s)\ ,\label{drag_E}
\end{equation}
where $\rho_w$, $c_w$, and ${\bf v}_w$ are the density, the sound speed and the bulk velocity of the wind, and ${\bf v}_s$ is the velocity of the solid body (Weidenschilling 1977; Shang 1998; Youdin \& Chiang 2004).  The gas drag depends on the density and velocity of the wind. The acceleration due to the gas drag is
\begin{equation}
{\bf a}_{\rm drag}\equiv\frac{{\bf F}_{\rm drag}}{m_s}=\frac{\rho_w }{\eta}[({\bf v}_w-{\bf v}_s)^2+c_w^2]^{1/2} ({\bf v}_w-{\bf v}_s)\ ,\label{drag_E_a}
\end{equation}
where we define the drag efficiency parameter $\eta\equiv R_s\rho_s$. The gas drag is more effective for a solid body with smaller $\eta$, meaning smaller or less dense.

Assuming that a solid body is launched at the initial angle of $\theta_0$ with respect to the midplane of the disk from the reconnection ring at the speed of the local X-wind, we integrate the equations of motion (\ref{EM1}) numerically to find the trajectory of the solid body. The allowing range of $\theta_0$ is $[0,\ \theta_w]$. In summary, the model calls 6 parameters: 3 parameter to specify the X-wind, $M_*$, $\dot{M}_{\rm D}$ and $\mu_*$; $f$ to specify the mass of the protoplanetary disk; $\eta$ to specify the efficiency of the gas drag; and $\theta_0$ to specify the initial launching angle of the solid body. For any particular state in Table \ref{Tab1}, the former 3 wind-related parameters are fixed, and the latter 3 parameters are free to be explored.

\section{Results}

\subsection{Trajectories of Solid Bodies}

\begin{figure}
\begin{center}
 \includegraphics[width=0.9\textwidth]{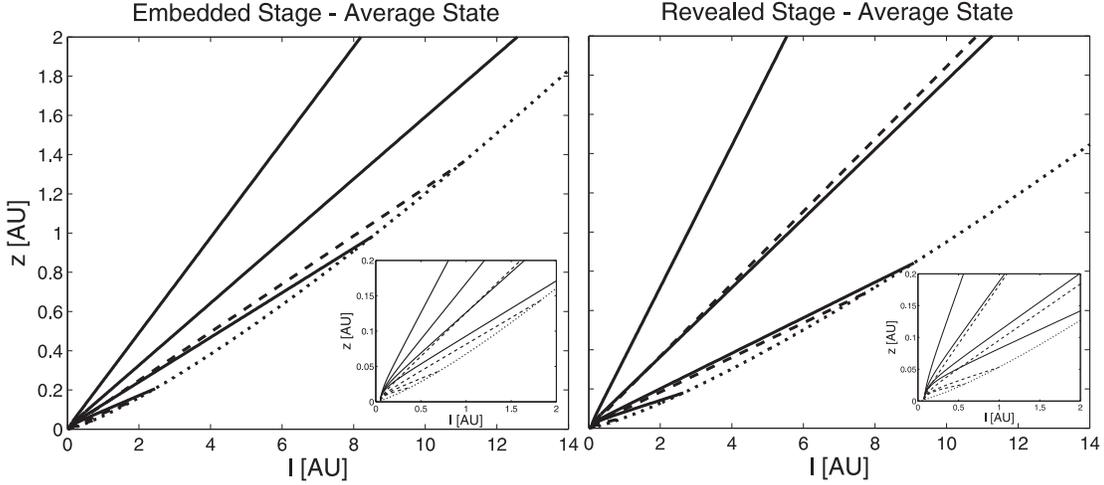}
 \caption{Trajectories of solid bodies entrained in X-winds of EA state (left) and RA state (right) with $f=5$. In both panels the solid lines are the trajectories of maximum initial launching angle ($\theta_0=\theta_w$), and the dashed lines are the trajectories of intermediate initial launching angle ($\theta_0=\theta_w/2$). The dotted lines show the scale height of the disk. Both panels have the same vertical axis, so the tags are only shown in the left panel. In the down-right region of each panel is the enlarged figure showing the detail of trajectories in $l<2$ AU.
 {\it Left}: From top to bottom the solid lines (or dashed lines) correspond to trajectories of $\eta\sim$ 0.6, 1.0, 1.3 and 1.6 cgs unit, respectively. Note that for $\theta_0=\theta_w$, the solid bodies of $\eta\sim$ 0.6 and 1.0 cgs unit are expelled from the planetary system; and for $\theta_0=\theta_w/2$, none of these solid bodies is expelled from the planetary system.
 {\it Right}: From top to bottom the solid lines (or dashed lines) correspond to trajectories of $\eta\sim$ 0.01, 0.03, 0.05 and 0.06 cgs unit, respectively. Note that for $\theta_0=\theta_w$, the solid bodies of $\eta\sim$ 0.01 and 0.03 cgs unit are expelled from the planetary system; and for $\theta_0=\theta_w/2$, only the solid body of $\eta\sim$ 0.01 cgs unit is expelled from the planetary system.}
 \label{Fig1}
  \end{center}
\end{figure}

The trajectories of solid bodies of different size and density (or $\eta$) in the EA and RA states are plotted in Figure \ref{Fig1}. Here we assume $f=5$ and only explore the situations of maximum lift ($\theta_0=\theta_w$) and intermediate lift ($\theta_0=\theta_w/2$). It is evident from Figure \ref{Fig1} that smaller and less dense objects can be launched to a larger orbit, in agreement with previous works (e.g. Shu et al. 1996). Note that although the left and the right panels of Figure \ref{Fig1} appear to be very similar, they show trajectories of solid bodies with very different $\eta$ (see the caption for details). Based on Figure \ref{Fig1}, the X-wind in the EA state can lift much larger and denser (or larger $\eta$) solid bodies to a significantly large radius than that in the RA state does.

The vertical extent of the disk intercepts some trajectories. From Figure \ref{Fig1}, the trajectories approach straight lines in the $l-z$ plot at large $l$ and $z$, in agreement with previous works of Shu et al. (1996), Shang (1998) and Liffman (2005). Without the consideration of disk's flared vertical geometry, these straight-line trajectories may easily be interpreted as `ejection'. For example, in the EA case, $f=5$ and $\theta_0=\theta_w$, the solid body with $\eta=1.3$ cgs unit re-enters the disk at $l=8.5$ AU (see the left panel of Figure \ref{Fig1}). If the disk was assumed to be a thin surface, such object would be ejected from the system. Therefore, one has to consider the disk's vertical geometry in order to yield realistic diagnostics on whether the solid bodies re-enter the disk or at what radius the solid bodies re-enter the disk.

The initial launching angle $\theta_0$ has a large effect on the trajectories of the solid bodies. From Figure \ref{Fig1}, a solid body is launched to a much larger orbit when $\theta_0=\theta_w$. For example, in the EA case and $f=5$, a solid body with $\eta=1$ cgs unit is ejected from the system if $\theta_0=\theta_w$, but re-enters the disk at $\sim2$ AU if $\theta_0=\theta_w/2$ (see Figure \ref{Fig1}). For another example, in the EA case and $f=5$, a solid body with $\eta=1.3$ cgs unit re-enters the disk at $\sim8.5$ AU if $\theta_0=\theta_w$, but re-enters the disk at much smaller radius ($\sim0.7$ AU) if $\theta_0=\theta_w/2$. Therefore, if numbers of solid bodies of such $\eta$ are initially launched from various angles in $[0,\ \theta_w]$, the X-wind spreads them to the disk's various annuli up to $\sim$8.5 AU.  In general, under the same condition, a solid body can be lifted to a larger orbit and thus re-enters the disk at a larger radius if initially launched from a larger angle.

There exists a threshold value of $\eta$ of solid bodies for any particular X-wind configuration, disk's mass and initial launching angle, denoted as $\eta_t$, below which the solid bodies will be expelled from the planetary system (see Figure \ref{Fig1}). This result is expected because the aerodynamic drag on the light particles by the wind is strong, i.e., these particles are more strongly coupled to the gas. For the heavy ($\eta>\eta_t$) particles, on the other hand, drag forces are much less important compared to the gravitational forces from the central star and the disk; and as a result, they fall back into the disk.

\subsection{The Effect of Disk Gravity}

In this work we include the gravity of the protoplanetary disk into the calculation of trajectories of solid bodies entrained in X-winds for the first time. Here we explore the effect of the disk gravity, and demonstrate that the disk mass modifies the trajectories and makes the re-entering radius lower significantly.

\begin{figure}
\begin{center}
 \includegraphics[width=0.5\textwidth]{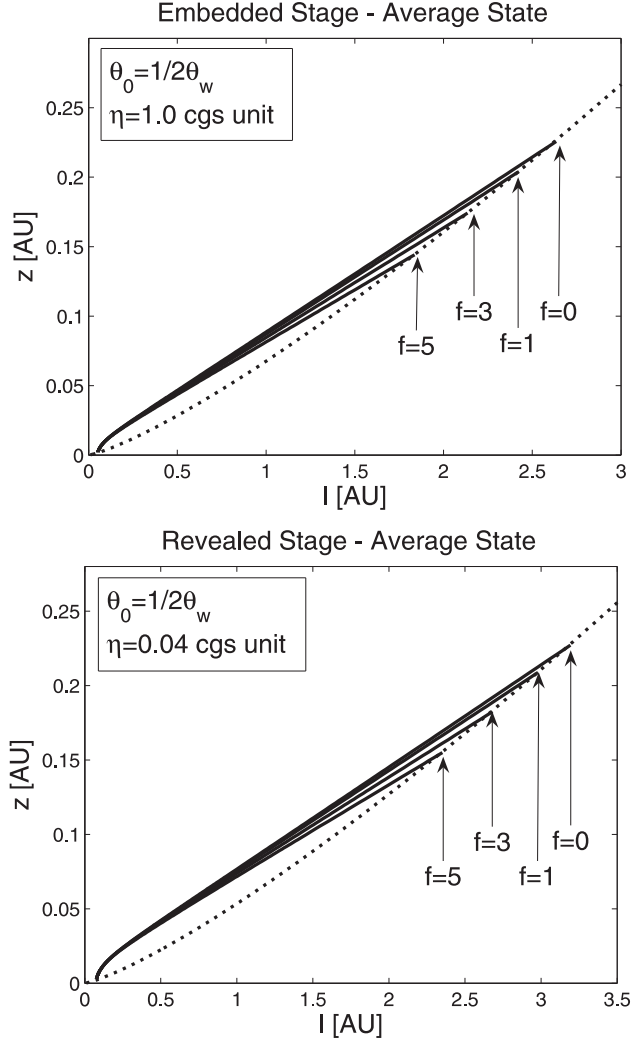}
 \caption{
 {\it Top}: Trajectories of a solid body of $\eta=1.0$ cgs unit entrained in the X-wind in the EA state with $\theta_0=\theta_w/2$.
  {\it Bottom}: Trajectories of a solid body of $\eta=0.04$ cgs unit entrained in the X-wind in the RA state with $\theta_0=\theta_w/2$.
  In both panels the trajectories correspond to the disk mass factor $f=0$, 1, 3, 5, respectively.
  The dotted lines show the scale height of the disk.
  Note that the vertical and horizontal scales in these panels are different.
 }
 \label{Fig2}
  \end{center}
\end{figure}

In Figure \ref{Fig2} we plot the trajectories of the same solid body with different disk mass factor $f$. If $f=0$, the disk's gravity is not considered and we return to the situation of Shu et al. (1996). From Figure \ref{Fig2}, we find that the disk's gravity has a negligible effect on the trajectory at $l<1$ AU. Detail analysis on the forces reveals that at $l<1$ AU, the gravity of the central star provides the major force in the $-{\bf e}_z$ direction. When $l>1$ AU, the trajectories under different $f$ become different. At such distance, the gravity of the protoplanetary disk becomes comparable with the gravity of the central star in the $-{\bf e}_z$ direction. As a result, the solid body flies closer to the disk (smaller $z$ at the same $l$) when the disk is more massive (larger $f$). Therefore, the re-entering radius becomes also smaller when the disk is more massive. Compared the MMSN disk case to the non-disk case, the re-entering radius decreases by $\sim10$\%. If still considered the fact that actual protoplanetary disk is very likely to be much more massive than the MMSN disk, for example $f=5$, the re-entering radius can be lowered by up to $\sim30$\%.

Here we find that the protoplanetary disk provides a non-negligible force in the $-{\bf e}_z$ direction and modifies the trajectories at $l>1$ AU. The previous calculations of Shu et al. (1996) that did not consider the disk gravity are therefore imperfect in estimating the re-entering radius.

\subsection{The Retention of Solids}

\begin{figure}
\begin{center}
 \includegraphics[width=0.5\textwidth]{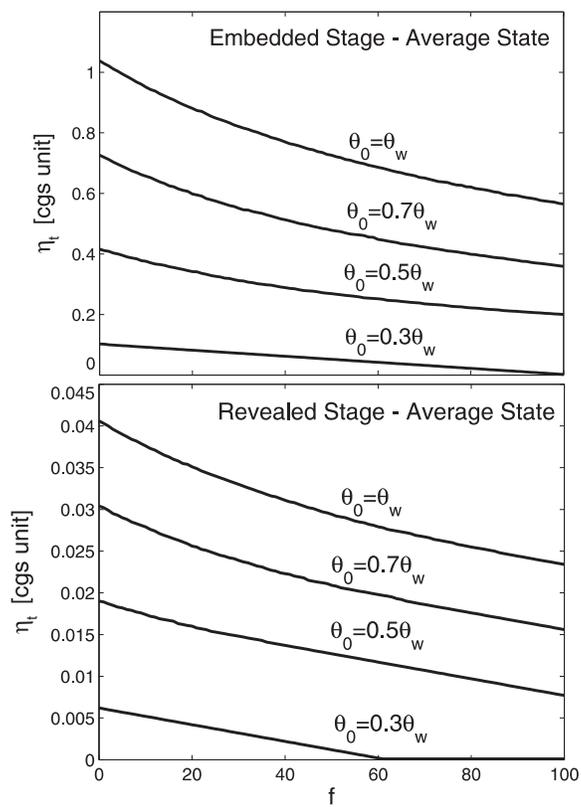}
 \caption{
 The sensitivity of the threshold $\eta_t$ on $f$ and $\theta_0$ in the EA state (top) and the RA state (bottom).
  For each line, we fix the value of $\theta_0$ as shown on the figure, vary $f$ and find the corresponding $\eta_t$.
 }
 \label{Fig2a}
  \end{center}
\end{figure}

In this section we study the sensitivity of $\eta_t$ on the disk's mass $f$ and the initial launching angle $\theta_0$. We show how $\eta_t$ depends on $f$ and $\theta_0$ in Figure \ref{Fig2a}. By definition, solid bodies of $\eta<\eta_t$ entrained in X-winds are expelled from the Solar system. As mentioned in Section 2.2, we consider a solid body to have been expelled if $l>40$ AU in the trajectory calculation. Note that the value of $\eta_t$ will change if this cutoff changes.

From Figure \ref{Fig2a}, the threshold $\eta_t$ depends most sensitively on the initial launching angle $\theta_0$. Typically when $f<10$, $\eta_t$ at the maximum launching angle $\theta_0=\theta_w$ is one-order-of-magnitude greater than that at a low launching angle of $\theta_0=0.3\theta_w$. For example, during the EA state and $f=5$, $\eta_t\sim0.98$ cgs unit at $\theta_0=\theta_w$; and $\eta_t\sim0.09$ cgs unit at $\theta_0=0.3\theta_w$. Similarly, during the RA state and $f=5$, $\eta_t\sim0.039$ cgs unit at $\theta_0=\theta_w$; and $\eta_t\sim0.006$ cgs unit at $\theta_0=0.3\theta_w$.

When the disk becomes more massive, the solids need to be smaller and lighter in order to be expelled by the X-winds. As seen in Figure \ref{Fig2a}, $\eta_t$ decreases by $\sim$10\% when $f$ increases from 0 to 10, and $\eta_t$ decreases by $\sim$30\% when $f$ increases from 0 to 100. The dependence of $\eta_t$ on $f$ is less sensitive, but should also be taken into account when considering the retention of solids in the protoplanetary disk.

The X-wind is stronger during the embedded stage, as a result, $\eta_t$ during the embedded stage is much larger than that during the revealed stage, as shown in Figure \ref{Fig2a}. It is interesting to note that $\eta_t$ during the revealed stage is compatible with the typical chondrule size range in meteorites, if the density is assumed to be $2\sim4$ g cm$^{-3}$. Finally, an important observation that we can make from Figure $\ref{Fig1}$ and Figure \ref{Fig2a} is that the horizontal distance that a solid body can travel depends on its size and density ($\eta$) very sensitively.

\subsection{The Size-Sorting Effect}

\begin{figure}
\begin{center}
 \includegraphics[width=0.5\textwidth]{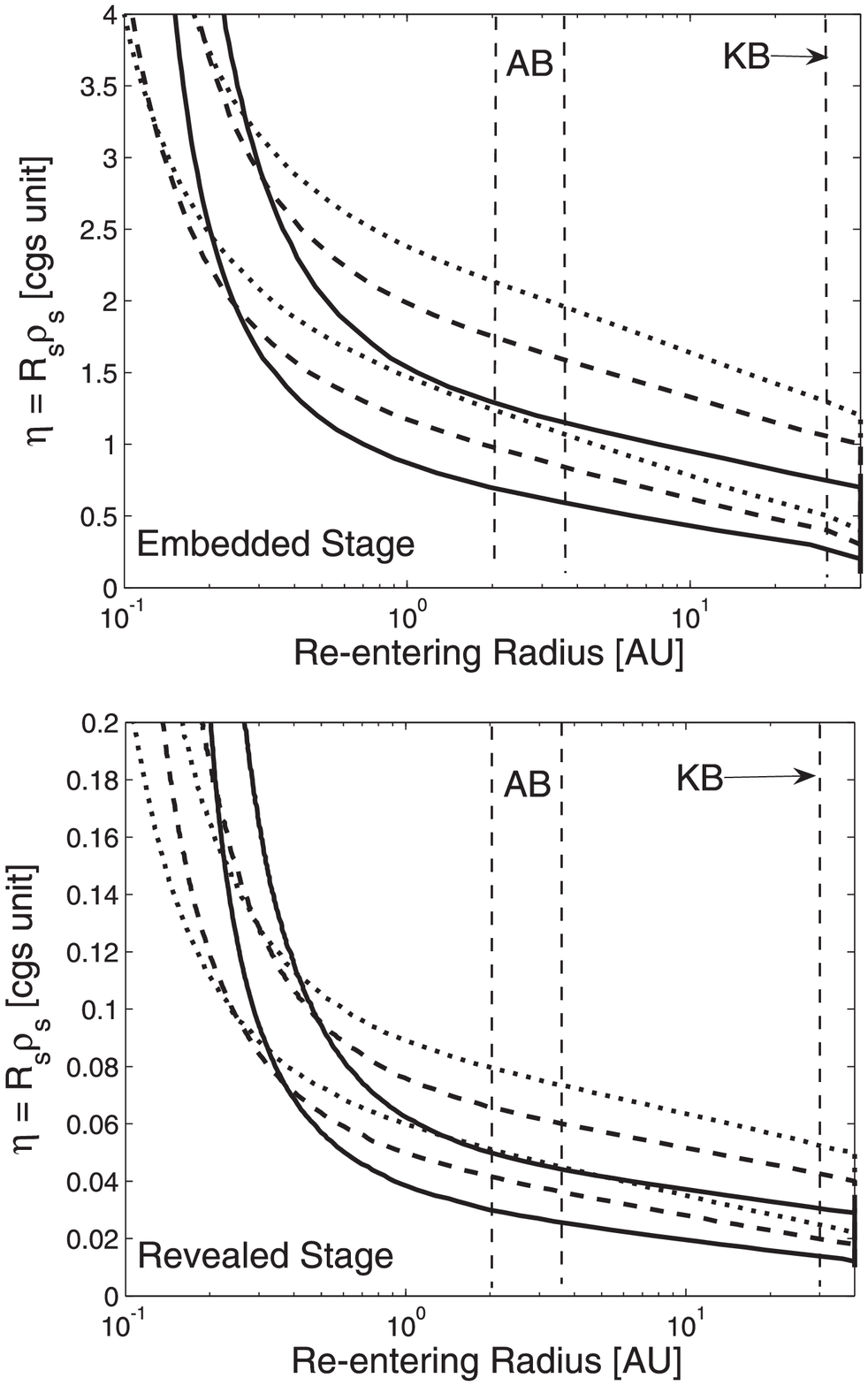}
 \caption{Relations between the solid body size and density ($\eta$) and the re-entering radius. Vertical light dashed lines indicates the radii of the asteroid belt (AB) and the Kuiper belt (KB) of the Solar system. For both stages, the solid curves describe the high state, the dashed curves describe the average state, and the dotted curves describe the low state. Each state is described with two curves. The higher curve corresponds to $f=0$ and $\theta_0=\theta_w$, and the lower curve corresponds to $f=5$ and $\theta_0=\theta_w/2$.}
 \label{Fig3}
  \end{center}
\end{figure}

In this section we investigate how the re-entering radius depends on the stellar magnetic dipole moment and size and density of the solid bodies. We calculate the trajectories of solid bodies of different $\eta$ in the states listed in Table \ref{Tab1}, and plot the relations between $\eta$ and the re-entering radius for each state in Figure \ref{Fig3}. There are still flexibility in choosing the value of parameter $f$ and $\theta_0$. In a particular state, a solid body can be launched to the maximum re-entering radius when $f=0$ and $\theta_0=\theta_w$, as plotted in Figure \ref{Fig3}. There is no theoretical minimum of the re-entering radius for any particular state, since $\theta_0$ can be as low as 0. However, we still plot the re-entering radius with $f=5$ and $\theta_0=\theta_w/2$ on Figure \ref{Fig3}, to indicate how small the re-entering radius can be for the intermediate lift of any particular state. Again, it is evident that smaller bodies re-enter the disk at larger radii.

The main result that one can draw from Figure \ref{Fig3} is that only the solid bodies whose $\eta$ are larger than but very close to $\eta_t$ of a particular state can be launched to the annuli of asteroid belt or Kuiper belt. For example, during the EH state and $f=0,\ \theta_0=\theta_w$, $\eta_t\sim$ 0.7 cgs unit, and the solid bodies with $\eta\sim$ 1.1 - 1.3 cgs unit re-enter the disk at the asteroid-belt radius (see the solid curve of the top panel of Figure \ref{Fig3}). For another example, during the RH state and $f=0,\ \theta_0=\theta_w$, $\eta_t\sim$ 0.03 cgs unit, and the solid bodies with $\eta\sim$ 0.04 - 0.05 cgs unit re-enter the disk at the asteroid-belt radius (see the solid curve of the bottom panel of Figure \ref{Fig3}). The densities of chondrules and CAIs are in a narrow range around 2-4 g cm$^{-3}$ (e.g. Hudges 1980). As a result, $\eta_t$ is in general equivalent to the threshold size. Of course, meteoritic parent bodies may have a wider feeding zone than the current asteroid belt location during their formation. However, the result does not change even if we consider a wider feeding zone, because in order to be lifted to a radius larger than 1 AU, the solid body must already have a size close to the threshold size (see Figure \ref{Fig3}).

In any particular state, the location of the asteroid belt, where we envisage the parent bodies of carbonaceous chondrites, can only receive solid bodies with sizes larger than but very close to the threshold size. However, the variation of initial launching angle $\theta_0$ affects $\eta_t$ significantly (see Figure \ref{Fig2a}) and enlarges the acceptable range of $\eta$ significantly. In reality, we know very little about the distribution of $\theta_0$. The X-wind model (Shu et al. 2001) suggests that the magneto-rotational instability (MRI) happens at the X-point to transport the rocks from the reconnection ring to the launching height. Based on the turbulent nature of the process, it is reasonable to assume that the distribution of $\theta_0$ is very wide, or nearly uniform. Imagine the CAIs and chondrules form in the X-region and are launched in X-wind from a wide range of $\theta_0$. Due to this variation of $\theta_0$, in any particular state, the annuli of the asteroid belt can receive a wide spectrum of solid bodies with different $\eta$, in other words, different sizes. For example, during the EH state, if the condition ($f=0,\ \theta_0=\theta_w$) is specified, the range for $\eta$ to re-enter the asteroid-belt radius is 1.1$\sim$1.3 cgs unit (see the solid curves of the top panel of Figure \ref{Fig3}). However, once the variation of $\theta_0$ from $\theta_w$ to $\theta_w/2$ is considered, the range for $\eta$ to re-enter the asteroid-belt radius becomes as wide as 0.6$\sim$1.3 cgs unit. In the previous section we find that $\eta_t$ depends most sensitively on $\theta_0$, so here the enlargement of the $\eta$ range is mainly due to the the variation in $\theta_0$.

Liffman (2005) predicts very effective size sorting and very sharp size distributions in a constant outflow. This is because in Liffman (2005), the initial angle of articles is fixed, corresponding to a single curve in our Figure \ref{Fig3}. Here we step forward to suggest that the variation of $\theta_0$ may make the size distribution of CAIs or chondrules in chondrites wider.

The second result is that an annulus of the protoplanetary disk receives one-order-of-magnitude smaller solid bodies (assuming the density is generally the same) during the revealed stage than during the embedded stage. For example the location of asteroid belt receives solid bodies of $\eta\sim$ 0.8-1.7 cgs unit during the EA state, but solid bodies of $\eta\sim$ 0.04-0.07 cgs unit during the RA state, with the uncertainty of $f$ and $\theta_0$ ($\sim[\theta_w/2,\theta_w]$) considered. Table \ref{Tab1} shows that only the EA and EL states can reach the required temperature $T_{\rm peak}>1800$ K and produce CAIs. In contrary, chondrules need lower temperature and are younger than CAIs (e.g. Scott 2007). It is then likely that chondrules mainly formed during the revealed stage. If this is the case, the chondrules should be smaller than CAIs in any particular chondritic group because of the X-wind transport. However, such difference could not be found in chondrites (e.g. Scott 2007). We will address to this discrepancy in the next section.

Last but not least, the variations of re-entering radius according to $\mu_*$ are different for small and large bodies. As the star becomes less magnetized, the small solid body re-enters the disk at a larger radius, but the large solid body re-enters the disk at a smaller radius. This feature may counter the physical intuition. Actually, as $\mu_*$ decreases, or the star becomes less magnetized, the accretion disk is truncated at a smaller radius. However, the star rotates much faster according to equation (\ref{Omega}) so that the terminal velocity of X-wind becomes faster. For small bodies, the wind terminal velocity determines how much momentum it can gain during the aerodynamic launching phase; therefore, less magnetized star produces faster winds, and then solid bodies re-enter the disk at larger radii. For large bodies, their aerodynamic launching phase is very short and their re-entering radii are close to the truncation radius $R_{\rm X}$.

\section{Discussion}

In any particular state of X-winds, $f$, $\theta_0$ and $\eta$ determine whether an entrained solid body re-enters the disk, as well as the re-entering radius. Section 3.3 shows that the retention of solid bodies depends most sensitively on the initial launching angle $\theta_0$. Especially, small bodies fall back to the disk if they are launched at a sufficiently small launching angle. In order to estimate the percentage of entrained solid bodies that eventually fall back, one has to know the distribution of $\theta_0$, which relates to complicated processes of retrieving materials from the reconnection ring.

Although blurred by the uncertainty in $\theta_0$, the size sorting effect of X-winds may have significant implications in our understanding of CAI and chondrule populations. Because of the uncertainty in $\theta_0$, the present model can only provide an upper limit of $\eta$ below which a solid body can be delivered by X-winds to certain location of the disk.

Based on Figure \ref{Fig3}, the acceptable $\eta$ for a solid body to be delivered to the location of the asteroid belt during the revealed stage is up to $0.08$ cgs unit. If we assume chondrules mainly formed during the revealed stage and the density of chondrules is in the range of 2$\sim$4 g cm$^{-3}$ (e.g. Hudges 1980), the X-wind transport model here predicts the maximum chondrule radius to be $\sim0.04$ cm, compatible with all carbonaceous chondrite groups except the CV class (e.g. Wurm \& Krauss, 2006). In average, CV chondrites have largest chondrules with size of $\sim0.1$ cm. From Figure \ref{Fig3}, chondrules in CV chondrites cannot be delivered to the location of the asteroid belt during the revealed stage. Instead, they might form during the EH state, or during the transitional stage between the embedded stage and the revealed stage.

In particular, CH chondrites have smallest chondrules with size of $\sim0.002$ cm (Jones et al. 2000). According to our X-wind transport model, a solid body as small as the chondrules in CH chondrites can easily be expelled from the system. One possible way to retain such tiny objects is to have very small initial launching angle $\theta_0$. However, if the chondrules that formed in the X-region were launched isotropically at all angles in $[0,\ \theta_w]$, large ones should be delivered simultaneously to the location of the asteroid belt. Therefore chondrules need to be launched preferably at small launching angles in order to explain the depletion of large chondrules in CH chondrites. Another possible explanation is that the chondrules of CH chondrites were delivered at the very end of the revealed stage, when the disk accretion and the X-wind approached to the end. If this is the case, chondrules in CH chondrite experienced the thermal treatment of a lower temperature, which could be tested by mineralogical examinations.

CAI-type and chondrulelike particles found in a short-term comet 81P/Wild 2 are very small having size of $\sim20$ $\mu$m (e.g. Brownlee et al. 2006; Nakamura et al. 2008). X-winds could easily expel these tiny objects from the Solar system (see Figure \ref{Fig3}). However, the particles had been larger than the present size because they were ablated and disaggregated during the deceleration in the aerogel (Flynn et al. 2006). The entrance-hole sizes of impact tracks place an upper limit for the size of the initial incoming particles, which is about 0.1 cm (or $\eta\sim0.3$ cgs unit). From Figure \ref{Fig3}, the solid bodies with such sizes can be transported to the Kuiper belt during the embedded stages.

CAI sizes are also different in various chondritic groups (Hezel et al. 2008), correlating with the chondrule sizes. CV chondrites appear to have largest CAIs with size range in 0.01$\sim$0.15 cm, and CH chondrites have smallest CAIs with size generally $<$0.01 cm (e.g. Zhang \& Hsu 2009). Because CAIs formation requires high temperature, only the EA and EL states are suitable (see dotted and dashed curve on the top panel of Figure \ref{Fig3}). We find that CAIs in all chondritic groups are lighter than the threshold $\eta_t$ of $\theta_0=\theta_w/2$ during the EA and EL states. Therefore, for these tiny primitive grains to be retained in the disk, the current X-wind model requires that they are launched preferably at low initial launching angles ($\theta_0<\theta_w/2$). This requirement provides a new aspect to test the X-wind model, and also suggests the importance to understand the initial $\theta_0$ distribution.  

Finally, according to Figure \ref{Fig3} large solid bodies whose $\eta$ well exceed $\eta_t$ cannot be lifted over long distances. They will re-enter the inner disk and follow the accreting inflow. It is possible that they enter the reconnection ring for the second time, during the timescale of embedded or revealed stage. After being thermally processed for the second time, they are expelled again by X-winds and then may enter a larger orbit if they experience some fragmentation. Especially, independent enveloping compound chondrules have a secondary layer of chondrulelike material enclosing an unrelated primary, indicating that they experience multiple melting (Wasson et al. 1995). It would be very interesting to study the role of X-wind in the multiple thermal treatments of such chondrules.

\section{Conclusion}

In this work we calculate the trajectories of solid body entrained in X-winds with the consideration of the protoplanetary disk's gravity and geometry. We find that the disk gravity becomes comparable to the gravity of the central star at $l>1$ AU, and the disk's mass affects the trajectories and the location where the solid bodies re-enter the disk significantly. The gravity of a disk whose mass is 5 times of the MMSN disk can lower the re-entering radius by up to 30\%.

The product of the size and the density, denoted as $\eta$ in this paper, is the key parameter to evaluate the efficiency the gas drag and the sorting effect. We find a threshold value of $\eta$, denoted as $\eta_t$, for any particular X-wind state, disk's mass and initial launching angle. Solid bodies with $\eta<\eta_t$ are expelled by the X-wind from the Solar system. We find that $\eta_t$ depends on the initial launching angle $\theta_0$ most sensitively, but also depends on the disk's mass factor $f$.

We demonstrate again the size-sorting effect of the X-wind transport by showing that only the solid bodies with a $\eta$ larger than but very close to $\eta_t$ can be launched to the location of the asteroid belt. We also find that the variation of the initial launching angle may smooth out the size-sorting effect, because solid bodies with a wider range of $\eta$ can be delivered to the desired location. This improved X-wind transport model is consistent with the chondrule size in all carbonaceous chondrite groups other than CV and CH. The size of chondrules in CV and CH chondrites, as well as the size of CAIs in carbonaceous chondrites, require additional mechanisms to be explained.

In order to compare the model prediction with the measured chondrule size distribution, it is necessary to further calculate the $\eta$ distribution after the X-wind size-sorting. To do so, one has to know the initial size distribution as well as the distribution over the initial launching angle, which are determined by the formation mechanism of chondrules in the X-region. Furthermore, there is little constrain on the variation of the stellar magnetic dipole moment. The final $\eta$ distribution reflects the entire evolution history of the early Sun. For example, if the stellar activity remains stable for a long period of time, we should see relatively sharp size distributions of chondrules. Therefore, the $\eta$ distribution of chondrules may contain valuable information of the early Solar System.

\acknowledgments

R. Hu is grateful to S. Seager for helpful advice in the preparation of the manuscript, to L. Elkins-Tanton for the encouragement of publication of this work, to J. Meyer and B. Weiss for helpful discussions, and to the anonymous referee for the improvement of the manuscript.

\clearpage

\begin{table*}
\caption{Definition and X-wind properties of all states considered in this paper. Each state is defined by a set of parameters ($M_*$, $\dot{M}_{\rm D}$, $\mu_*$). For each case, key X-wind properties $R_{\rm X}$, $\Omega_{\rm X}$, $\bar{v}_w$ and $T_{\rm peak}$ are calculated based on equations (\ref{RX}-\ref{Tpeak}). Full names of cases are given following the table, but simplified names, given in the first column of the table, are used in the text.}
\begin{tabular}{llccccccc}
  \hline
  States$^1$ & $M_*$ & $\dot{M}_{\rm D}$ & $\mu_*$ & $R_{\rm X}$ & $\Omega_{\rm X}$ & $\bar{v}_w$ & $T_{\rm peak}$$^2$ \\
  & ($M_{\odot}$) & (M$_{\odot}$ yr$^{-1}$) & (G cm$^{-3}$) & (cm) & (s$^{-1}$) & (cm s$^{-1}$) & (K) \\
  \hline
  EH & 0.5 & $2\times10^{-6}$ & 6    & $1.43\times10^{12}$ & $4.76\times10^{-6}$ & $1.43\times10^7$ & 1325 \\
  EA & 0.5 & $2\times10^{-6}$ & 2    & $7.64\times10^{11}$ & $1.22\times10^{-5}$ & $1.96\times10^7$ & 1837 \\
  EL & 0.5 & $2\times10^{-6}$ & 1    & $5.14\times10^{11}$ & $2.21\times10^{-5}$ & $2.39\times10^7$ & 2270 \\
  \hline
  RH & 0.8 & $1\times10^{-7}$ & 3    & $2.12\times10^{12}$ & $3.34\times10^{-6}$ & $1.49\times10^7$ & 941 \\
  RA & 0.8 & $1\times10^{-7}$ & 1    & $1.13\times10^{12}$ & $8.57\times10^{-6}$ & $2.04\times10^7$ & 1299 \\
  RL & 0.8 & $1\times10^{-7}$ & 0.5  & $7.61\times10^{11}$ & $1.55\times10^{-5}$ & $2.48\times10^7$ & 1598 \\
  \hline
\end{tabular}
\label{Tab1}
 \begin{enumerate}
 \item[1]
 \begin{itemize}
 \item EH --- Embedded Stage, High State
 \item EA --- Embedded Stage, Average State
 \item EL --- Embedded Stage, Low State
 \item RH --- Revealed Stage, High State
 \item RA --- Revealed Stage, Average State
 \item RL --- Revealed Stage, Low State
 \end{itemize}
 \item[2] For EH, EA and EL cases, $L_*=4.4L_{\odot}$ and $R_*=3R_{\odot}$ are used in the calculation of $T_{\rm peak}$, where $L_{\odot}$ and $R_{\odot}$ is the current luminosity and radius of the Sun. Similarly, for RH, RA and RL cases, we assume $L_*=2.5L_{\odot}$ and $R_*=3R_{\odot}$.
 \end{enumerate}
\end{table*}

\clearpage

\clearpage

\clearpage

\clearpage

\begin{thebibliography}{}

\bibitem[]{} Brownlee, D. et al. 2006, Science, 314, 1711

\bibitem[]{} Ciesla, F. J., 2009, Icarus, 44, 1663

\bibitem[]{} Clayton, R. N. 1993, Annu. Rev. Earth Planet. Sci., 21, 115

\bibitem[]{} Clayton, R. N. 2002, \nat, 415, 860

\bibitem[]{} Clayton, R. N., Mayeda, T. K., Goswami, J. N., Olsen, E. J. 1991, Geochimica et Cosmochimica Acta, 55, 2317

\bibitem[]{} Desch, S. J. 2007, \apj, 671, 878

\bibitem[]{} Desch, S. J., Morris, M. A., Connolly, H. C., 2010, 41st Lunar and Planetary Science Conference, 2200

\bibitem[]{} Flynn, G. J., et al. 2006, Science, 314, 1731


\bibitem[]{} Hartmann, L. 2009, Accretion processes in star formation, 2nd edition, Cambridge University Press

\bibitem[]{} Hayashi, C. 1981, Prog. Theor. Phys. Suppl., 70, 35

\bibitem[]{} Hezel, D. C., Russel, S. S., Ross, A. J., Kearsley, A. T. 2008, Meteoritics \& Planetary Science, 43, 1879

\bibitem[]{} Hudges, D. W. 1980, Earth \& Planetary Science Letters, 51, 26-28

\bibitem[]{} Jones, R. H., Lee, T., Connolly, H. C., Love, S. G., Shang, H. 2000, Formation
of chondrules and CAIs: Theory vs. observation. In: Mannings, V., Boss,
A.P., Russell, S.S. (Eds.), Protostars and Planets, vol. IV. Univ. of Arizona
Press, Tucson, pp. 927–962.

\bibitem[]{} Liffman, K. 2005, Meteoritic \& Planetary Science, 40, 123

\bibitem[]{} Nakamura, T. et al. 2008, Science, 321, 1664

\bibitem[]{} Rubin, A. E. 2000, Earth Science Review, 50, 3

\bibitem[]{} Rubin, A. E. 2005, Geochimica et Cosmochimica Acta, 69, 4907

\bibitem[]{} Scott, E. R. D. 2007, Annu. Rev. Earth Planet. Sci., 35, 577

\bibitem[]{} Shang, H., Allen, A., Li, Z.-Y., Liu, C.-F., Chou, M.-Y., Anderson, J., 2006, \apj, 649, 845

\bibitem[]{} Shang, H. 1998. Protostellar Winds, Jets, and Chondritic Meteorites (Doctoral Dissertation).

\bibitem[]{} Shu, F. H., Najita, J., Ostriker, E., Wilkin, F., Ruden, S., Lizano, S. 1994, \apj, 429, 781

\bibitem[]{} Shu, F. H., Najita, J., Ruden, S., Lizano, S. 1994, \apj, 429, 797

\bibitem[]{} Shu, F. H., Shang, H., Lee, T. 1996, Science, 271, 1545

\bibitem[]{} Shu, F. H., Shang, H., Glassgold, A. E., Lee, T. 1997, Science, 277, 1475

\bibitem[]{} Shu, F. H., Shang, H., Gounelle, M., Glassgold, A. E., Lee, T. 2001, \apj, 548, 1029

\bibitem[]{} Swindle, T. D., Campins, H. 2004, Meteoritics \& Planetary Sci., 39, 1733

\bibitem[]{} Tayler, S. R. 1992, Solar System Evolution (Cambridge University Press, Cambridge, UK)

\bibitem[]{} Wasson, J. T., Krot, A. N., Lee, M. S., Rubin, A. E. 1995, Geochim. Cosmochim. Acta., 59, 1847

\bibitem[]{} Weisberg, M. K., McCoy, T. J., Krot, A. N. 2006, Systematics and Evaluation of Meteorite Classification. In, Meteorites and the Early Solar System II, 19-52 (D.S. Lauretta and H.Y. McSween, Eds.), Univ. Arizona press

\bibitem[]{} Weidenschilling S. J. 1977, \mnras, 180, 57

\bibitem[]{} Wood, J. A. 2004, Geochimica et Cosmochimica Acta, 68, 4007

\bibitem[]{} Wooden, D. H. 2008, Space Science Review, 138, 75

\bibitem[]{} Wurm, G., Krauss O. 2006, Icarus, 180, 487

\bibitem[]{} Wood, J. A. 2004, Geochimica et Cosmochimica Acta, 68, 4007

\bibitem[]{} Youdin A. N., Chiang E. I. 2004, \apj, 601, 1109

\bibitem[]{} Zhang, A., Hsu, W. 2009, Meteoritics \& Planetary Science, 44, 787

\end{thebibliography}
\end{document}